\documentclass[11pt,dvipdfmx]{article}

\renewcommand{\thefootnote}{\fnsymbol{footnote}}

\usepackage{amsbsy,amssymb,latexsym,amsfonts,amsmath}
\usepackage{mathrsfs}
\usepackage[pdftex]{graphicx}
\usepackage{bm}
\usepackage{here}

\numberwithin{equation}{section}
\newcommand{\bel}[1]{\begin{equation}\label{#1}}                     
\newcommand{\bal}[1]{\begin{eqnarray}\label{#1}}                     
\newcommand{\be}{\begin{equation}}
\newcommand{\ee}{\end{equation}}
\newcommand{\im}{\mathrm{i}}
\newcommand{\ex}{\mathrm{e}}
\newcommand{\de}{\mathrm{d}}
\newcommand{\dis}{\displaystyle}

\newcommand{\qq}{\qquad}

\begin{document}
%
%
\begin{titlepage}
\begin{flushright}
\normalsize
~~~~
OCU-PHYS 555\\
December, 2021\\
\end{flushright}

\vspace{15pt}

\begin{center}
{\huge  Perturbation of multi-critical unitary matrix models,
 } \\
\vspace{15pt}
{\huge double scaling limits, and Argyres-Douglas theories}\\
\end{center}

\vspace{43pt}

\begin{center}
{Takeshi Oota$ $\footnote{e-mail: toota@osaka-cu.ac.jp}}\\

%
\vspace{10pt}
%

\it Osaka City University Advanced Mathematical Institute (OCAMI)

\vspace{5pt}

3-3-138, Sugimoto, Sumiyoshi-ku, Osaka, 558-8585, Japan \\

\end{center}
%
\vspace{15pt}
\begin{center}
Abstract\\
\end{center}
Using the saddle point method, 
we give an explicit form of the planar free energy and Wilson loops
of unitary matrix models in the one-cut regime.
The multi-critical unitary matrix models are shown to undergo
third-order phase transitions at two points by studying the planar free energy. 
One of these ungapped/gapped phase transitions is multi-critical,
while the other is not multi-critical.
The spectral curve of the $k$-th multi-critical matrix model exhibits an $A_{4k-1}$ singularity
at the multi-critical point.
Perturbation around the multi-critical point and its double scaling limit are studied.
In order to take the double scaling limit, 
the perturbed coupling constants should be fine-tuned
such that all the zero points of the spectral curve approach to the $A_{4k-1}$ singular point.
The fine-tuning is examined in the one-cut regime, 
and the scaling behavior of the perturbed couplings is determined.
It is shown that the double scaling limit of the spectral curve is isomorphic to the Seiberg-Witten curve
of the Argyres-Douglas theory of type $(A_1, A_{4k-1})$.


\vfill

\end{titlepage}

\renewcommand{\thefootnote}{\arabic{footnote}}
\setcounter{footnote}{0}

\section{Introduction}

Unitary one matrix models have been studied from various viewpoints: as a toy model of two-dimensional QCD \cite{GW80,wad1212,wad80},
in relation to the string theories \cite{KMS0309,pol9703}, 
as exactly solvable quantum theories \cite{man90,PS90a,PS90b,MP90,DT90,HP90,min91,CM91,ake9806},
and in connection with integrable systems and their hierarchies \cite{BMS91,CDM91,wat91,KM92,ABI91,ABS9112,AB9203,DJMW9206,LM9206,his9609,his9611,his9705}.
For reviews of the unitary matrix models, see, for example, \cite{RCV9609,mor0906,mor15,mar1206}.

Recently, the unitary matrix models have attracted renewed interest in the context of matrix model/gauge theory correspondence \cite{IOY1805,IOY1812,IOY1909,MOT1909,IY2103,KZ2105}. (See also \cite{MMM0909,IMO0911,EM0911,SW0911,SW0911,MMS0911,IO1003,MMM1003,MS1004,EM1006,IOY1008,MY1009,MMS1011,BMTY1011,NR1112,GMM1205,bou1206,NR1207} for correspondence between 
more general matrix models and supersymmetric gauge theories.)
The unitary model with a potential of degree $k$ corresponds to the $\widehat{A}_{2k,2k}$ theory \cite{CV1103,BMT1112}.
Its $k$-th multi-critical point \cite{man90,PS90b} is related 
to the type $(A_1, A_{4k-1})$ Argyres-Douglas (AD) superconformal fixed point \cite{AD,APSW}
of the supersymmetric theory.
In \cite{IOY1805,IOY1812,IOY1909,IY2103}, it is shown that the limit to the AD point of the gauge theory
can be described by the double scaling limit to the multi-critical point in the unitary matrix model\footnote{
Double scaling limits of $\mathcal{N}=1$ supersymmetric gauge theories to Argyres-Douglas type singularities are studied in \cite{fer0211,BHM0603}.}.

In \cite{IY2103}, perturbations around the $k$-th critical point and its double scaling limit
are studied by using the method of orthogonal polynomials \cite{bes79,IZ80,PS90a,PS90b}.
Using the planar string equations, the scaling operators are determined. Their scaling dimensions agree with
those of the $(A_1, A_{4k-1})$ AD theory.

In this paper, we reconsider these perturbations by using the saddle point method \cite{BIPZ78,GW80,wad1212}.
Before considering the perturbation, we study large-$N$ properties of the unperturbed unitary matrix model
with the multi-critical coupling constants. 

First, we consider a one-cut solution of general unitary matrix model.
The eigenvalue density, planar Wilson loops and planar free energy are explicitly calculated. 
In the one-cut regime, i.e. in the one-gap phase, the spectral curve for the unitary model is explicitly determined in terms of the coupling constants.

Secondly, we consider the unitary matrix model with the multi-critical coupling constants.
This model depends only on one real parameter $g$, the inverse of the ’t Hooft coupling. 
It turns out that
the eigenvalue distribution develops at most one gap.
We will see that the multi-critical model has three phases.
There are two values of $g$ at which the ungapped/gapped transition occurs.
One corresponds to the $k$-th multi-critical point, while the other to an ordinary critical point.
The $k=1$ case is the famous Gross-Witten-Wadia (GWW) unitary matrix model \cite{GW80,wad1212,wad80}.
The ungapped/gapped phase transition is known to be third-order.
The models for $k>1$ are a generalization of the GWW model. 
We show that all these phase transitions are also third-order.

Finally, we consider perturbation of the multi-critical matrix model and its double scaling limit.
To take a double scaling limit, the perturbed coupling constants should be fine-tuned such that 
all the zero points of the matrix model spectral curve approach to the critical point.
Then the double scaling limit of the spectral curve is isomorphic 
to the Seiberg-Witten (SW) curve of the $(A_1, A_{4k-1})$ AD theory.

The fine-tuning of the coupling constants is examined in the one-cut regime.
The scaling behavior of the perturbed couplings is determined. 
The obtained perturbed potential is consistent with the form of the scaling operators constructed in \cite{IY2103}.
The analysis of \cite{IY2103} shows that the dimensions of the scaling operators agree with those of the parameters of the SW curve.  
Here, we determine more explicitly how the perturbed couplings of the unitary model
are related to the parameters in the SW curve. The double scaling limit of
the spectral curve in the one-cut regime is degenerated. 

To resolve the degeneracy, the double scaling limit should be taken in the multi-cut regime with $2k$ cuts.
In this case, filling fractions $\nu_i=N_i/N$ must be assigned to extrema of the potential. Here $N=N_1+N_2+\dotsm+N_{2k}$. 

This paper is organized as follows. 
In section 2, we briefly review the saddle point method for the case of
unitary one matrix model with general coupling constants. We then give the one-cut solution to the planar resolvent and
the eigenvalue density.
In section 3, we calculate the explicit form of the Wilson loops and the free energy at the planar level in the one-gap phase.
In section 4, we report properties of the unitary matrix model with multi-critical coupling constants. 
The model has one ungapped phase and two gapped phases. The planar Wilson loops and the planar free energy
in these three phases are determined. 
In section 5, the double scaling limits
of the spectral curve are investigated.
In section 6, some discussions are given.

\section{Unitary matrix model with general couplings}

\subsection{Partition function and planar free energy}

In order to explain our notation,  
we briefly review the unitary one matrix model with general coupling constants.

The partition function of the unitary matrix model can be written as
\bel{ZG1}
Z(N, \{t\}) = \frac{1}{\mathrm{vol}(U(N))}
\int[ \de U ] \, \exp\left(  N \mathrm{Tr}\, W(U) \right),
\ee
where the potential $W(z)$ is given by
\bel{Wz}
 W(z) = \sum_{n \geq 1} \frac{1}{n} \bigl( t_n\, z^n + t_{-n}\, z^{-n} \bigr).
\ee
Here $\{ t \} = \{ t_n \}_{n \neq 0}$ denotes the set of coupling constants.
The integration over the $U(N)$ unitary matrices is reduced to the eigenvalue integrals:
\bel{ZG2}
Z(N, \{t\}) = \frac{1}{N!} \int \frac{\de^N \theta}{(2\pi)^N}
\left[ \prod_{1 \leq i<j \leq N} 2 \sin\left( \frac{\theta_i - \theta_j}{2} \right) \right]^2
\exp\left( - N \sum_{i=1}^N V(\theta_i) \right),
\ee
where $V(\theta)=-W(\ex^{\im \theta})$. We assume that the potential $V(\theta)$ is real, i.e. 
$t_n^* = t_{-n}$.

The planar free energy is given by
\be
F_0(\{t\}) = \lim_{N \rightarrow \infty} \frac{1}{N^2} \ln Z(N,\{t\}).
\ee

\subsection{Saddle point method}

In the saddle point approach, the planar free energy is expressed by using the
eigenvalue density $\rho(\theta)$ as follows:
\bel{PFEG}
F_0 =-  \int \de \theta \rho(\theta) V(\theta)
+ \mathrm{P} \int \de \theta \, \rho(\theta) \int \de \theta'\, \rho(\theta')
\ln \left| 2 \, \sin \left( \frac{\theta - \theta'}{2} \right) \right|.
\ee
The density function $\rho(\theta)$ is obtained by solving the saddle point equation
\bel{SPEG}
V'(\theta) = \mathrm{P} \int \de \theta'\, \rho(\theta') \cot\left(\frac{\theta-\theta'}{2} \right),
\qq ( \theta \in \mathrm{supp}\,  \rho )
\ee
with the normalization condition
\bel{RNC}
\int \de \theta\, \rho(\theta) = 1.
\ee
Using a solution to the saddle point equation, the planar free energy is given by
\bel{F0rho}
F_0 = - \frac{1}{2} \int \de \theta \, \rho(\theta) V(\theta)
- \frac{1}{2} V(\theta_0) + \mathrm{P} \int \de \theta\, \rho(\theta)
\ln \left| 2 \sin \left( \frac{\theta_0 - \theta}{2} \right) \right|,
\ee
where $\theta_0$ is an arbitrary point in the support of the density function. The last two terms in the right-handed side of \eqref{F0rho}
come from the Lagrange multiplier for the normalization condition \eqref{RNC}.

\subsection{Planar resolvent and spectral curve}

The planar resolvent is defined by
\be
\omega(z):=\lim_{N \rightarrow \infty} 
\left\langle \frac{1}{N} \mathrm{Tr}\, \frac{1}{z-U} \right\rangle
= \sum_{n=0}^{\infty} \frac{\mathcal{W}_n}{z^{n+1}},
\ee
where $\mathcal{W}_n$ are the planar part of the expectation values of the Wilson loop operators.
\be
\mathcal{W}_n = \lim_{N \rightarrow \infty} 
\left\langle \frac{1}{N} \mathrm{Tr}\, U^n \right\rangle.
\ee
Let $\widetilde{W}(z) = W(z) - \ln z$.
Then, ignoring an irrelevant phase factor $(-1)^{(1/2)N(N-1)}$, the partition function \eqref{ZG2} can be rewritten in the form similar to that of the Hermitian
matrix model
\be
Z = \frac{1}{N!} \int \frac{\de^N z}{(2\pi)^N}
\prod_{i<j} (z_i-z_j)^2 \, \exp\left(  N \sum_{i} \widetilde{W}(z) \right).
\ee
The planar resolvent satisfies the loop equation
\be
\omega(z)^2 + \widetilde{W}'(z)\,  \omega(z) - f(z) = 0,
\ee
where
\be
f(z) = \lim_{N \rightarrow \infty} 
\left\langle \frac{1}{N} \mathrm{Tr} \left(\frac{\widetilde{W}'(z) - \widetilde{W}'(U)}{z- U}\right)
\right\rangle.
\ee
By solving the loop equation, the resolvent is determined as
\bel{PR}
\omega(z) = -\frac{1}{2} \widetilde{W}'(z) + \frac{1}{2} \sqrt{ \widetilde{W}'(z)^2 + 4\, f(z)}.
\ee
By introducing $y:=2\, \omega(z) + \widetilde{W}'(z)$, we get the spectral curve for the unitary matrix model:
\be
y^2 = \widetilde{W}'(z)^2 + 4\, f(z).
\ee

\subsection{One-cut solution}

Multi-cut solutions of the unitary matrix models \cite{GW80,man90} are obtained by 
generalizing the method for polynomial potentials \cite{BIPZ78,mig83}
to the case of Laurent polynomials. We remark that there is another approach \cite{miz0411}
which uses a mapping from a Hermitian matrix model and
the technique in the Hermitian matrix model to construct the multi-cut solutions.
In this paper, we take the first approach and consider the one-cut solution in detail.

A one-cut solution for the planar resolvent is obtained by setting the square root in \eqref{PR} as
\be
\sqrt{ \widetilde{W}'(z)^2 + 4\, f(z)}= \frac{1}{z} M(z) \sqrt{\sigma(z)},
\qq
\sigma(z) = (z-\alpha)(z-\beta),
\ee
where the constants $\alpha$ and $\beta$ denote the endpoints of the cut in the $z$-plane.
Here $M(z)$ is a Laurent polynomial in $z$.  It is determined as
\bel{Mz}
M(z) = -\int_{C} \frac{\de w}{2\pi \im} \frac{ z \, \widetilde{W}'(z) - w \, \widetilde{W}'(w)}{(z-w) \sqrt{\sigma(w)}},
\ee
where $C$ is a contour which encircles the cut counterclockwise. The endpoints of the cut, $\alpha$ and $\beta$, 
are determined by the conditions
\bel{EQab}
\int_C \frac{\de w}{2\pi \im} \frac{\widetilde{W}'(w)}{\sqrt{\sigma(w)}} = 0,
\qq
\int_C \frac{\de w}{2\pi \im} \frac{w\, \widetilde{W}'(w)}{\sqrt{\sigma(w)}} = 2.
\ee
The planar resolvent \eqref{PR} in the one-gap phase can be written as
\bel{PR2}
\omega(z) = \frac{1}{2z} \int_C \frac{\de w}{2\pi \im} \frac{w \, \widetilde{W}'(w)}{(z-w)}
\sqrt{\frac{\sigma(z)}{\sigma(w)}}.
\ee
The second equation of \eqref{EQab} is equivalent to the condition that 
the resolvent $\omega(z)$ behaves as $\omega(z) = 1/z + O(1/z^2)$ for large $z$.

The eigenvalue density $\rho(\theta)$ in the one-gap phase can be read off from the discontinuity of
the resolvent $\omega(z)$ along the cut:
\be
\rho(\theta) = \frac{1}{2 \pi} M(\ex^{\im \theta}) \sqrt{\sigma(\ex^{\im \theta})}.
\ee

\section{Unitary matrix model in one-gap phase}

For the one-cut solution, the contour integrals are easily performed and explicit expressions of various quantities
can be obtained.

It is convenient to set
\be
x:= \frac{\alpha+\beta}{2\, \sqrt{\alpha \beta}}, \qq
v:=\sqrt{\alpha \beta}.
\ee
Then $\sigma(z) = z^2 -2\, v x  z + v^2$ and $M(z)$ \eqref{Mz} is evaluated as
\bel{Mz2}
M(z) = \sum_{n \geq 1} t_n 
\left( \sum_{m=0}^{n-1} v^m
z^{n-1-m}\, P_m(x) \right) +
\sum_{n \geq 1} t_{-n} \left(
\sum_{m=0}^{n-1} v^{-m-1} z^{-n+m} P_m(x) \right),
\ee
where $P_n(x)$ is the Legendre polynomial.
The conditions \eqref{EQab} can be written as
\be
1 = \sum_{n \geq 1} t_n\, v^n \, P_{n-1}(x) - \sum_{n \geq 1} t_{-n}
\, v^{-n} P_n(x),
\ee
\be
1 = \sum_{n \geq 1} t_{-n}\, v^{-n} P_{n-1}(x) - \sum_{n \geq 1} t_n \, v^n\, P_n(x).
\ee
We also use the parameterization:
\be
x = \cos \theta_c, \qq \xi:= \frac{1+x}{2} = \cos^2(\theta_c/2),
\qq
v = \ex^{\im \varphi}, \qq
\alpha = \ex^{\im( \varphi - \theta_c)}, \qq
\beta = \ex^{\im ( \varphi + \theta_c)}.
\ee
The density function in the one-gap phase is given by
\bel{rhoOC}
\rho(\theta) = \frac{1}{\pi} \widetilde{M}(\theta)
\sqrt{ \sin^2 \left(\frac{\theta_c}{2} \right) - \sin^2 \left( \frac{\theta - \varphi}{2} \right) },
\ee
for $(\varphi - \theta_c) \leq \theta \leq (\varphi+\theta_c)$ and $\rho(\theta)=0$ otherwise.
Here
\be
\begin{split}
\widetilde{M}(\theta) &= \sum_{n \geq 1} ( t_n\, v^n + t_{-n}\, v^{-n})
\sum_{m=1}^{n} P_{n-m}(x) \cos \left( m - \frac{1}{2} \right) ( \theta -\varphi) \cr
& + \im \sum_{n \geq 1} ( t_n\, v^n - t_{-n} \, v^{-n} )
\sum_{m=1}^{n} P_{n-m}(x) \sin \left( m - \frac{1}{2} \right) ( \theta- \varphi).
\end{split}
\ee
We have checked that \eqref{rhoOC} satisfies the normalization condition \eqref{RNC}.
In the $\theta_c \rightarrow \pi$ limit, the gap in the eigenvalue distribution vanishes and
the density function \eqref{rhoOC} goes to the solution in the ungapped phase \cite{GW80,man90}:
\bel{grho0}
\lim_{\theta_c \rightarrow \pi} \rho(\theta) = \frac{1}{2\pi}
\left[ 1 + \sum_{n \geq 1} ( t_n\, \ex^{\im n \theta} + t_{-n}\, \ex^{-\im n \theta} ) \right].
\ee

The planar Wilson loops in the one-gap phase can be calculated by using the density function \eqref{rhoOC} as
\be
\mathcal{W}_{n} = \int_{\varphi- \theta_c}^{\varphi + \theta_c} \de \theta\, \rho(\theta)\, 
\ex^{\im n \theta}, \qq (n \in \mathbb{Z}).
\ee
Note that $\mathcal{W}_0=1$. 
Let
\be
\mathscr{K}_n(x):=\frac{1}{\pi} \int_{-\theta_c}^{\theta_c} 
\de \theta\, 
\cos\left( n + \frac{1}{2} \right) \theta\,
\sqrt{\sin^2 \left( \frac{\theta_c}{2} \right) - \sin^2 \left( \frac{\theta}{2} \right)}, \qq
(n \in \mathbb{Z}).
\ee
It satisfies
\be
\mathscr{K}_{-n}(x) = \mathscr{K}_{n-1}(x),
\ee
and 
\be
\mathscr{K}_n(x) = \frac{1}{n+1} \left( \frac{1-x}{2} \right) P_n^{(1,-1)}(x), \qq (n \geq 0)
\ee
where $P^{(\alpha, \beta)}_n(x)$ is the Jacobi polynomial.

For $n \geq 1$, we obtain
\be
\mathcal{W}_{\pm n}
= \sum_{m \geq 1} m\, t_{\pm m}\, v^{\pm(m+n)}\, \mathscr{P}_{m,n}(x)
+ \sum_{m \geq 1} m\, t_{\mp m}\, v^{\mp(m-n)}\, \mathscr{Q}_{m,n}(x),
\ee
where for $m, n \geq 1$, 
\be
\mathscr{P}_{m,n}(x):= \frac{1}{m} \sum_{r=1}^m P_{m-r}(x)\, \mathscr{K}_{n+r-1}(x),
\ee
\be
\mathscr{Q}_{m,n}(x):=\frac{1}{m} \sum_{r=1}^m P_{m-r}(x)\, \mathscr{K}_{n-r}(x).
\ee
We find that 
these $\mathscr{P}_{m,n}(x)$ and $\mathscr{Q}_{m,n}(x)$ are symmetric 
under exchange of their indices: $\mathscr{P}_{n,m}(x) = \mathscr{P}_{m,n}(x)$,
$\mathscr{Q}_{n,m}(x) = \mathscr{Q}_{m,n}(x)$. They can be rewritten as
\be
\mathscr{P}_{m,n}(x) = \frac{(1-x)(1+x)}{4(m+n)}
\Bigl[ P^{(1,0)}_{m-1}(x) P^{(0,1)}_{n-1}(x) + P^{(0,1)}_{m-1}(x) P^{(1,0)}_{n-1}(x) \Bigr],
\ee
\be
\mathscr{Q}_{m,n}(x) = \frac{(1-x)}{2\, m} \delta_{m,n}
+ \frac{(1-x)(1+x)}{4\, \mathrm{min}(m,n)}
\sum_{r=|m-n|+1+\delta_{m,n}}^{\mathrm{max}(m,n)}
\frac{1}{(r-1)} P_{r-|m-n|-1}(x) P^{(1,1)}_{r-2}(x).
\ee
In particular, for $n=1$, the Wilson loop $\mathcal{W}_1$ is given by
\be
\begin{split}
\mathcal{W}_1 &=\frac{1}{2}(1-x) t_{-1} + \frac{1}{4}(1-x)(1+x)
\sum_{m \geq 1} t_m\, v^{m+1} P_{m-1}^{(1,1)}(x) \cr
& + \frac{1}{4}(1-x)(1+x) \sum_{m \geq 2} \left( \frac{m}{m-1} \right)
t_{-m}\, v^{-m+1}\, P_{m-2}^{(1,1)}(x).
\end{split}
\ee

By setting $\theta_0=\varphi$ in \eqref{F0rho}, we obtain the planar free energy in the one-gap phase:
\be
\begin{split}
F_0 &=\sum_{n \geq 1} \frac{1}{2 n} (t_n\, \mathcal{W}_n + t_{-n} \, \mathcal{W}_{-n} )
+ \sum_{n \geq 1} \frac{1}{2n} (t_n\, v^n + t_{-n}\, v^{-n} ) \cr
&+\sum_{n \geq 1} ( t_n\, v^n + t_{-n}\, v^n )
\left[ \sum_{m=1}^n \mathcal{H}_{m-1}(x) P_{n-m}(x) \right] + \frac{1}{2} \ln \left( \frac{1-x}{2} \right),
\end{split}
\ee
where
\be
\mathcal{H}_n(x):=\frac{1}{2} \sum_{r=0}^n (-1)^r \frac{(n+r)!}{r!\, (n-r)!\, (r+1)!}
\Bigl[ \psi(r+1/2) - \psi(r+2) + 2\, \ln 2 \Bigr] \left( \frac{1-x}{2} \right)^{r+1}.
\ee
Here $\psi(z) = \Gamma'(z)/\Gamma(z)$ is the digamma function.
Note that $(1/2) \ln ((1-x)/2) = \ln( \sin(\theta_c/2))$.
See Appendix A for some integral formula.

\subsection{Remarks on the case of symmetric couplings}

For the case of symmetric coupling constants, $t_{-n} = t_n$, we have $v^2=1$
and $x$ is a solution of
\bel{XCS}
1 = \sum_{n \geq 1} t_n \, v^n (P_{n-1}(x) - P_n(x) ).
\ee
In this case,  the Wilson loops take the form
\be
\mathcal{W}_n = \mathcal{W}_{-n} 
= \left[ 1 - \left( \frac{1+x}{2} \right)^2 \mathscr{S}_{2n-2}(x) \right] t_n
+ \frac{1}{2} (1-x)(1+x)^2 \sum_{m \neq n} m\, t_m \, v^{m+n}\,
\mathscr{T}_{m-1,n-1}(x).
\ee
where $\mathscr{S}_{2n-2}(x)$ and $\mathscr{T}_{m-1,n-1}(x)$ are certain polynomials
of degree $2n-2$ and degree $m+n-3$, respectively.
For $n=1$, the Wilson loop $\mathcal{W}_1$ is given by
\be
\mathcal{W}_1
= \left[ 1 - \left( \frac{1+x}{2} \right)^2 \right] t_1
+ \frac{1}{4}(1-x)(1+x)^2 \sum_{m \geq 2} 
\left( \frac{m}{m-1} \right) t_m\, v^{m-1} \, P^{(1,2)}_{m-2}(x).
\ee

The eigenvalue density $\rho(\theta)$ for the cubic and quartic potentials in $\cos \theta$
is obtained in \cite{man90} and \cite{PS90a}, respectively. Our density \eqref{rhoOC} 
and the condition \eqref{XCS} with $v=1$ 
reproduce their results by specifying couplings to their values\footnote{If
the factor $(1+k_1+2k_2+3k_3)$ in the equation for $a$
\cite[p.1329]{PS90a} is replaced by $(k_1+2k_2+3k_3)$.}.

The free energy $F_0$ in the one-gap phase for the quadratic cosine potential is calculated in \cite{man90}.
There is a slight discrepancy between our result and eq.(15) of \cite{man90}:
$F_0 - F_0^{(\mathrm{Mandal})} = (3/\lambda) b^2\, a_2$.

\section{Unitary matrix model with multi-critical couplings}

The unitary matrix model with the $k$-th critical potential is 
a case of the symmetric couplings $t_{-n} = t_n$.
The $k$-th multi-critical couplings are determined for $k \leq 5$ in \cite{man90} and
for general $k$ in \cite{PS90b}. They are given by
\be
t_n = t_{-n} = g \, t^{(k)}_n,
\qq
t^{(k)}_n = \frac{\Gamma(k+1)^2}{\Gamma(k+n+1)\Gamma(k-n+1)}.
\ee
Note that $t^{(k)}_n=0$ for $n>k$. Here $g$ is the inverse of the ’t Hooft coupling. 
Since we have assumed that the potential is real, $g$ is a real parameter.
We will see that there are three phases in the large-$N$ limit: one ungapped and two gapped phases.
The eigenvalue distribution is gapless for $g_*<g<1$,
while it develops one gap for $g>1$ or for $g<g_*$. Here $g_*$ is a negative constant (see \eqref{gstar}).
We call the gapped phase with $g>1$ (resp. $g<g_*$)  \textit{phase Ia} (resp. \textit{phase Ib}).

Changing the normalization of the potential by $W(z) \rightarrow g W(z)$,
$V(\theta) \rightarrow g V(\theta)$,
the partition function of the $k$-th critical model is written as
\bel{Ztheta}
\begin{split}
Z(N, g) &= \frac{1}{\mathrm{vol}(U(N))}
\int [ \de U ] \, \exp\Bigl(  N\, g\, \mathrm{Tr}\, W(U) \Bigr) \cr
&=\frac{1}{N!}
\int \frac{\de^N \theta}{(2\pi)^N}
\left[ \prod_{1 \leq i<j \leq N} 2\, \sin \left( \frac{\theta_i - \theta_j}{2} \right) \right]^2
\exp\left( -N \, g \sum_{i=1}^N V(\theta_i) \right),
\end{split}
\ee
where
\bel{potW}
W(z) = \sum_{n=1}^k \frac{t^{(k)}_n}{n} \bigl( z^n + z^{-n}), \qq
V(\theta) = -W(\ex^{\im \theta})=- 2 \sum_{n=1}^k \frac{t^{(k)}_n}{n} \cos n \theta.
\ee
The planar free energy in this case is denoted by
\be
F_0(g) = \lim_{N \rightarrow \infty} \frac{1}{N^2} \ln Z(N, g).
\ee
The partition function \eqref{Ztheta}
is normalized such that $Z(N, 0) = 1$. Therefore, $F_0(0)=0$.

\subsection{Properties of the potential $V(\theta)$}

Some properties of the potential $V(\theta)$ \eqref{potW} are summarized here.

The derivative of the potential $V(\theta)$
\be
V'(\theta) = 2 \sum_{n=1}^k t^{(k)}_n \, \sin n \theta
\ee
vanishes only at $\theta=0, \pi$ (mod $2\, \pi$). Therefore, the eigenvalue distribution develops at most one gap in the large-$N$ limit.

The potential $W(z)$ \eqref{potW} is obtained from that of \cite{PS90b} by flipping the variable $z \rightarrow -z$,
or equivalently, by shifting the angle $\theta \rightarrow \theta +\pi$. Due to this shift, 
the minimum of the potential $V(\theta)$ is located at $\theta=0$: 
\bel{Vmin}
V_{\mathrm{min}}^{(k)} := V(0) = -\sum_{n=1}^k \frac{1}{n}=- \psi(k+1) - \gamma_{\mathrm{E}}.
\ee
where $\gamma_{\mathrm{E}}$
is the Euler's constant, 
and the maximum is at $\theta=\pi$:
\bel{Vmax}
V_{\mathrm{max}}^{(k)}:=V(\pi)=\sum_{n=1}^k \left[ \frac{1}{n-(1/2)} - \frac{1}{n} \right]
= \psi(k+(1/2)) - \psi(k+1) + 2\, \ln 2.
\ee

As a series in $k$, $V_{\mathrm{max}}^{(k)}$ is monotonically increasing and is bounded from above
\be
0<V_{\mathrm{max}}^{(k=1)} < V_{\mathrm{max}}^{(k=2)}< \dotsm < 2\, \ln 2.
\ee
The potential $V$ behaves near the minimum and near the maximum respectively as
\bel{VNmin}
V(\theta) = V(0) +\frac{k}{2} \theta^2 - \frac{k^2}{4!} \theta^4
+ \frac{k^2(2\, k-1)}{6!} \theta^6 +O(\theta^8),
\ee
\bel{VNmax}
V(\pi + \theta) = V(\pi) + \frac{k}{2(2\, k-1)} \theta^2
+ \frac{k^2}{4!\, (2\, k-1)(2\, k-3)} \theta^4 + O(\theta^6).
\ee

\subsection{Ungapped phase $(g_*<g<1)$}

The eigenvalue density in the ungapped phase is obtained by
substituting the multi-critical couplings into \eqref{grho0}:
\be
\rho(\theta) = \frac{1}{2\pi} \left( 1 + 2 \, g \sum_{n=1}^k t^{(k)}_n \, \cos n \theta \right).
\ee
The density must be non-negative for any real $\theta$.
This holds for $g_* \leq g \leq 1$ where
\bel{gstar}
\frac{1}{g_*}=1 - \frac{(1)_k}{(1/2)_k} < 0.
\ee
Here $(a)_k = \Gamma(a+k)/\Gamma(a)$.
First few values of $g_*$ are listed:
\be
g_*^{(k=1)} = -1, \qq
g_*^{(k=2)} = - \frac{3}{5}, \qq
g_*^{(k=3)} = - \frac{5}{11}, \qq
g_*^{(k=4)} = -\frac{35}{93}.
\ee

The planar free energy in the ungapped phase is determined as
\bel{F00}
F_0^{(\mathrm{ungapped})}(g) = \frac{1}{2} \, C_k\, g^2,
\ee
where
\be
C_k := 2 \sum_{n=1}^k \frac{(t^{(k)}_n)^2}{n} 
= \sum_{n=1}^k \left( \frac{1}{n} - \frac{1}{k+n} \right)
= 2 \, \psi(k+1) - \psi(2k+1) + \gamma_{\mathrm{E}}.
\ee

The Wilson loops in this phase are given by
\bel{ugWn}
\mathcal{W}_n = \mathcal{W}_{-n}
= \int \de \theta\, \rho(\theta) \cos n \theta =g \, t^{(k)}_n.
\ee

\subsection{Gapped phase (Phase Ia; $g>1$)}

In the phase Ia, we have $v=1$, i.e. $\varphi=0$. The equation for $x=\cos \theta_c$ \eqref{XCS} is given by
\be
1= g \bigl ( 1 - \xi^k  \bigr),
\ee
where $\xi=(1+x)/2$.
Here we have used
\be
\sum_{n=1}^k t^{(k)}_n \bigl( P_{n-1}(x) - P_n(x) \bigr) = 1 - \left( \frac{1+x}{2} \right)^k.
\ee  
Therefore, the edge of the gap is related to the ’t Hooft coupling as follows
\bel{kappa}
\xi = \cos^2 (\theta_c/2)= \left( 1 - \frac{1}{g} \right)^{1/k}.
\ee
For the case of GWW model ($k=1$), this is equivalent to the condition for $\theta_c$ \cite{GW80}:
\be
\sin^2(\theta_c/2) = \frac{1}{g}.
\ee
For $k \geq 2$, \eqref{kappa} reflects the fact that the critical point at $g=1$ is multi-critical with the exponent $\gamma=-1/k$. 

\subsubsection{Planar resolvent and Wilson loops}

In this case, the Laurent polynomial \eqref{Mz2} is simplified to
\be
M(z) = \frac{t^{(k)}_k \, g}{z^{k}} \left( \sum_{n=0}^{k-1}
\frac{\Gamma(2n+1)}{\Gamma(n+1)^2} (\xi z)^n (z+1)^{2k-2n-1} \right).
\ee
Using this expression, we find that the planar resolvent can be rewritten as
\bel{PRIa}
\omega(z)
= \frac{1}{z} + \sum_{n=1}^k \frac{t^{(k)}_n g}{z^{n+1}}
- \frac{(g-1)}{2\, z}
\left[
F\left(\frac{1}{2}, k, k+1; \frac{4\, \xi\, z}{(z+1)^2} \right) - 1 \right],
\ee
where $F(\alpha, \beta, \gamma; z)$ is the Gauss's hypergeometric function. 
This is suitable for examining the large $z$
expansion. For $n \geq 1$, we obtain the planar Wilson loops
\bel{Pn}
\mathcal{W}_n = g\, t^{(k)}_n 
- (g-1) \sum_{j=1}^n (-1)^{n-j} \left(\frac{k}{k+j} \right)
\frac{\Gamma(n+j)}{\Gamma(n-j+1)} \frac{\xi^{j}}{j!\, (j-1)!}.
\ee
Recall that $t^{(k)}_n=0$ for $n > k$. 
We remark that this is consistently extended to the result in the ungapped phase  \eqref{ugWn} 
by setting $\xi=0$.

\subsubsection{Eigenvalue density}

The eigenvalue density in the phase Ia is given by
\bel{rhogap}
\rho(\theta)
=\frac{2\, (k!)^2}{(2k)!\, \pi} \, g
\left( \sum_{n=0}^{k-1}
\frac{(2\, n)!}{(n!)^2} \xi^n \left( 4 \cos^2 \frac{\theta}{2} \right)^{k-n-1}
\right)
\cos \frac{\theta}{2} \sqrt{\sin^2 \frac{\theta_c}{2} - \sin^2 \frac{\theta}{2}}
\ee
for $-\theta_c \leq \theta \leq \theta_c$ and $\rho(\theta)=0$ otherwise.

\subsubsection{Planar free energy}

The planar free energy in the phase Ia is given by
\bel{F0}
F_0
=\frac{1}{2}\,  C_k\, g^2
+ \frac{1}{2} \, g\sum_{n=1}^k \left[ \frac{(2-g)}{n} + \frac{(g-1)}{k+n} \right] \xi^n
+ \frac{1}{2} \ln (1-\xi).
\ee
This expression is extendable to the ungapped phase \eqref{F00} by setting $\xi=0$, i.e.
\be
\xi = \begin{cases}
0,& (g_* < g<1), \cr
(1-1/g)^{1/k},& (g>1).
\end{cases}
\ee

We remark that $F_0$ can be also obtained by integrating its derivative:
\be
\frac{\partial F_0}{\partial g} = \sum_{n=1}^k \frac{2\, t^{(k)}_n}{n} \mathcal{W}_n
=g \sum_{n=1}^k \left( \frac{1}{n} - \frac{1}{k+n} \right)
(1-\xi^{k+n}).
\ee
In this case, the integration constant should be zero in order to make $F_0(g)$  continuous at $g=1$.

In the gapped phase Ia, \eqref{F0} can be rewritten as
\be
F_0^{(\mathrm{Ia})}
=\frac{1}{2} C_k\, g^2
+ \frac{1}{2} \, g\sum_{n=1}^{k-1} \left[ \frac{(2-g)}{n} + \frac{(g-1)}{k+n} \right] \xi^n
- \frac{1}{4\, k}(g-1)(g-3)
+ \frac{1}{2} \ln (1-\xi).
\ee

For $k=1$, \eqref{F0} reproduces the planar free energy of the GWW model:
\be
F_0^{(k=1)} =  \frac{1}{4}(1-\xi)g^2+\frac{3}{4}  \xi\, g
+\frac{1}{2} \ln(1-\xi) 
=
\begin{cases}
\dis \frac{1}{4} g^2, & (-1<g<1), \cr
 & \cr
\dis g - \frac{1}{2} \ln g - \frac{3}{4}, & (g>1).
\end{cases}
\ee
The free energy $F_0^{(k=1)}$ for $g<-1$ is given in \eqref{Fok1Ib}.

\subsubsection{Third-order phase transition at $g=1$ with $\gamma=-1/k$}

In this gapped phase, let $g=1+\epsilon$ ($\epsilon>0$). The planar free energy behaves for $\epsilon \rightarrow 0$
as
\bel{F0Iacrit}
F_0^{(\mathrm{Ia})} - \frac{1}{2}\, C_k \, g^2 =
-\frac{k^2}{(k+1)(2\, k+1)} \epsilon^{2+(1/k)} + O(\epsilon^{2+(2/k)}).
\ee
This leading behavior is consistent with the result obtained by the method of orthogonal polynomials \cite{IY2103}.
This equation implies that the exponent for the critical point $g=1$ is given by $\gamma=-1/k$.
The third derivative of the planar free energy is discontinuous at $g=1$.
Therefore, this gapped/ungapped phase transition is third-order for all $k$.

\subsubsection{Large $g$ limit}

As a consistency check, 
we study the large $g$ limit of the free energy $F_0(g)$  \eqref{F0}
and compare it with the result from direct integration of the partition function \eqref{Ztheta}.

For large $g \gg 1$, it is convenient to change the integration variables in \eqref{Ztheta}
from $\theta_i$ to $\lambda_i = \sqrt{g}\, C\, \theta_i$
where $C$ is a normalization constant. With \eqref{VNmin}, this change of variables gives
\be
-N \, g \sum_i V(\theta_i) = - N^2\, V(0)\, g
- \frac{Nk}{2\, C^2} \sum_i \lambda_i^2 + O(1/g).
\ee
The partition function \eqref{Ztheta} approaches to the that of the Hermitian matrix model with
the Gaussian potential. If we approximate the integration range $- \pi\sqrt{g}\, C \leq \lambda_i \leq \pi \sqrt{g}\, C$
by $-\infty < \lambda_i < \infty$,  we can perform the integration to find
\be
Z(N, g) =  \frac{G_2(N+1)}{(2\pi)^{(1/2)N}}
\left( \frac{1}{k\, g\, N} \right)^{(1/2)N^2} \ex^{-N^2 \, V(0) g} \left( 1 +O(1/g) \right),
\ee
which leads to
\bel{F0lg}
F_0 = \left( \sum_{n=1}^k \frac{1}{n} \right) g - \frac{1}{2} \ln (kg) - \frac{3}{4} +O(1/g).
\ee

Using \eqref{kappa}, the large $g$ behavior of \eqref{F0} is determined as
\be
\begin{split}
F_0 &= \left( \sum_{n=1}^k \frac{1}{n} \right) g
- \frac{1}{2} \ln(k g)
- \frac{3}{4}+ \frac{(k-1)}{12\, k\, g}
+ \frac{(k-1)(5\, k-1)}{288 \, k^2 \, g^2} 
+ \frac{(k-1)(3\, k-1)}{480\, k^2\, g^3} \cr
&+ \frac{(k-1)(251\, k^3-109\, k^2 + k+1)}{86400\, k^4 \, g^4}
+ \frac{(k-1)(5k-1)(19 \, k^2 - 6 \, k-1)}{60480\, k^4\, g^5}
+O(g^{-6}).
\end{split}
\ee
This is consistent with the Gaussian approximation \eqref{F0lg}.

Furthermore, by setting $\lambda=\theta/\theta_c$ with $-1 \leq \lambda \leq 1$,  
the eigenvalue density \eqref{rhogap} shows the Wigner's semi-circle law \cite{wig55,wig58}
in the large $g$ limit:
\be
\rho(\theta) \, \de \theta= \left( \frac{2}{\pi} \sqrt{1-\lambda^2} + O(1/g) \right) \de \lambda.
\ee
Note that $\theta_c \simeq 2/\sqrt{kg}$ for $g \gg 1$.

\subsection{Gapped phase (Phase Ib; $g<g_*$)}

For $g<0$, the minimum of the potential $g\, V(\theta)$ is located at $\theta=\pi$.
The eigenvalues distributes around $\theta=\pi$, and the gap of the eigenvalue density opens at $\theta=0$ for $g<g_*<0$. 

In the phase Ib, we have $v=-1$, i.e. $\varphi=\pi$. The equation for $x$ \eqref{XCS} becomes
\bel{XCB}
\frac{1}{g} =  \sum_{n=1}^k (-1)^n t^{(k)}_n \, ( P_{n-1}(x) - P_n(x) ).
\ee
This equation can be rewritten as
\be
\frac{1}{g} - \frac{1}{g_*}
= \sum_{n=1}^k (-1)^{n-1} \frac{(1/2)_n}{(1/2)_k} \frac{(k!)^2}{(n!)^2\, (k-n)!} \, \xi^n
= \frac{k\, (1)_k}{2\, (1/2)_k} \xi + O(\xi^2).
\ee
At $\xi=0$, $g$ takes the critical value $g_*$. 
This signals the phase transition at $g=g_*$ has the critical exponent $\gamma=-1$.
Around $\xi=0$,
the solution $\xi=\xi(g)$ must start as
\be
\xi = \frac{2}{k(1-g_*)} \left( 1 - \frac{g_*}{g} \right) + \dotsm.
\ee
Let
\be
\delta:= \frac{1}{k(1-g_*)} \left( 1 - \frac{g_*}{g} \right)
= \frac{(1/2)_k}{k\, (1)_k} \left( \frac{1}{g} - \frac{1}{g_*} \right).
\ee
Then, first few terms of the perturbative solution is given by
\be
\xi = 2\, \delta + \frac{3}{2} ( k-1) \, \delta^2
+ \frac{1}{12}(k-1) (17\, k^2 - 7) \, \delta^3 + 
\frac{5}{96} (k-1)(28\, k^2 - 17\, k + 3) \, \delta^4 + O(\delta^5).
\ee

The equation \eqref{XCB} can be also rewritten as
\bel{igb1}
-\frac{1}{g} = \sum_{n=1}^k \frac{k!}{n!\, (k-n)!} \frac{(1/2)_n\, (1/2)_{k-n}}{(1/2)_k}\, (1-\xi)^n
= \frac{k}{(2\, k-1)} (1-\xi) + O\bigl((1-\xi)^2 \bigr).
\ee
Note that $\xi \rightarrow 1$ for $g \rightarrow -\infty$.
Near $\xi=1$, we have
\be
\begin{split}
1-\xi &= \frac{(2\, k-1)}{k} (-g)^{-1}
- \frac{3(k-1)(2\, k-1)^2}{k^2\, (2\, k-3)}
(-g)^{-2} \cr
&+ \frac{(k-1)(2\, k-1)^3 (8\, k^2 - 28\, k +15)}{k^3 (2\, k-3)^2(2\, k-5)}
(-g)^{-3}
+O\bigl((-g)^{-4} \bigr).
\end{split}
\ee

\subsubsection{Planar resolvent and Wilson loops}

The Laurent polynomial $M(z)$ \eqref{Mz2} is given by
\be
M(z) = g \sum_{n=1}^k t^{(k)}_n \sum_{m=0}^{n-1} (-1)^m P_m(x) ( z^{n-1-m}  - z^{-n+m} ) .
\ee
The planar resolvent $\omega(z)$ is obtained from the general one-cut solution
by setting the couplings to the multi-critical values. 
No significant simplification is observed in the gapped phase Ib.

The Wilson loops take the form
\be
\mathcal{W}_n = g\, t^{(k)}_n - g\, \xi^2 \, \Delta \mathcal{W}_n,
\ee
where
\be
\Delta \mathcal{W}_n = \mathscr{S}_{2n-2}(x) - 2(1-x) \sum_{m \neq n} (-1)^{m+n}\,  m\, t^{(k)}_m
\mathscr{T}_{m-1,n-1}(x).
\ee
For example, the explicit form for $n=1,2$ is given by
\be
\Delta \mathcal{W}_1
= \sum_{r=0}^{k-1} (-1)^r \frac{(k!)^2}{r!\, (r+2)!\, (k-r-1)!} \frac{(1/2)_{r+1}}{(1/2)_k} \xi^r,
\ee
\be
\Delta \mathcal{W}_2
= \sum_{r=0}^k (-1)^r
\frac{(k!)^2}{r!\, (r+2)!\, (k-r)!}\frac{(1/2)_r}{(1/2)_k} 
\Bigl[ (2\, r+1) k + r(r+2) \Bigr] \xi^r.
\ee

\subsubsection{Eigenvalue density}

The eigenvalue density in the phase Ib is given by
\be
\begin{split}
\rho(\theta) 
&=\frac{2 (-g)}{\pi}
\left[ \sum_{n=1}^k \left( \sum_{m=n}^k (-1)^{m-n} \, t^{(k)}_m \, P_{m-n}(x) \right)
\sin \left( n - \frac{1}{2} \right) \theta
\right] \,  \cr
& \qq \times  \sqrt{
\sin^2 \left(\frac{\theta_c}{2}\right) - \sin^2 \left(\frac{\theta-\pi}{2} \right)
},
\end{split}
\ee
for $(\pi - \theta_c) \leq \theta \leq (\pi + \theta_c)$ and $\rho(\theta)=0$ otherwise.

We can check that this density in the large $(-g)$ limit
 is consistent with the Wigner's semi-circle law by using the change of variable
 $\theta = \pi + \theta_c\, \lambda$ ($-1 \leq \lambda \leq 1$).

\subsubsection{Planar free energy}

The planar free energy in the phase Ib takes the form
\bel{F0Ib}
F^{(\mathrm{Ib})}_0(g)
= g^2 \,f_{0,2}(x) + g\, f_{0,1}(x)  - \frac{g}{2} V(\pi) + \frac{1}{2} \ln(1-\xi),
\ee
where 
\be
f_{0,2}(x) :=\frac{1}{2} \sum_{m=1}^k \sum_{n=1}^k
(-1)^{m+n} \left( \frac{m}{n} + \frac{n}{m} \right)
t^{(k)}_m\, t^{(k)}_n\, (\mathscr{P}_{m,n}(x) + \mathscr{Q}_{m,n}(x) ),
\ee
\be
f_{0,1}(x):=2 \sum_{n=1}^k (-1)^n t^{(k)}_n \left[ \sum_{m=1}^n \mathscr{H}_{m-1}(x) P_{n-m}(x) \right].
\ee
Especially for $k=1$, we have
\be
F_0^{(\mathrm{Ib}, k=1)} = \frac{1}{4} g^2 (1-\xi^2) - \frac{1}{2} g\, \xi + \frac{1}{2} \ln (1-\xi),
\ee
and the condition \eqref{XCB} is given by
\be
1 = (-g) (1-\xi).
\ee
Using these relations, we find
\bel{Fok1Ib}
F_0^{(\mathrm{Ib}, k=1)} =(-g) - \frac{1}{2} \ln(-g) - \frac{3}{4}, \qq (g<-1).
\ee

\subsubsection{Third-order phase transition at $g=g_*$ with $\gamma=-1$}

Around $\xi=0$, the inverse of the ’t Hooft coupling $g$ approaches to its critical value $g_*$ as
\be
g = g_* - \frac{k}{2} g_*(g_*-1)\, \xi + g_*(g_*-1) 
\left[ \frac{k^2}{4}(g_*-1) + \frac{3}{16} k(k-1) \right] \xi^2 + \dotsm.
\ee
Then, the free energy $F_0$ behaves as
\be
F_0^{(\mathrm{Ib})} - \frac{1}{2} C_k \, g^2 = - \frac{k^2}{24}(1-g_*)^2 \, \xi^3 + O(\xi^4)
= - \frac{1}{3\, k\, g_*^3(g_*-1)} ( g_* - g)^3 + O\bigl( (g_*-g)^4 \bigr).
\ee
This shows that the gapped/ungapped phase transition at $g=g_*$ is also third-order but it is not multi-critical.
Its exponent is $\gamma=-1$.

\subsubsection{Large $(-g)$ limit}

Let us consider the  large $(-g)$ limit of the planar free energy \eqref{F0Ib}.
By changing variables from $\theta_i$ to $\lambda_i$ with  $\theta_i = \pi + \lambda_i/(\sqrt{-g}\,  C)$ for a constant $C$,
the Gaussian approximation of  the partition function  \eqref{Ztheta} yields
\be
Z(N,g) =
\frac{G_2(N+1)}{(2\pi)^{(1/2)N} }
\left[ \frac{(2\, k-1)}{N \, k \, (-g)} \right]^{(1/2)N^2}  \ex^{-N^2\, V(\pi) g }
\Bigl[ 1 + O(1/(-g)) \Bigr].
\ee
Here we have used \eqref{VNmax}.
From this, we find
\bel{GF0b}
F_0=V(\pi) \, (-g) - \frac{1}{2} \ln \left( \frac{k(-g)}{2\, k-1} \right) - \frac{3}{4} +O(1/(-g)).
\ee

On the other hand, 
by examining the large $(-g) \gg 1$ behavior of \eqref{F0Ib}, we obtain
\be
F_0(g) = V(\pi) \, (-g) - \frac{1}{2} \ln \left( \frac{k(-g)}{2\, k-1} \right) - \frac{3}{4}
- \frac{(k-1)(2\, k-1)}{4\, k (2\, k-3)} \left( - \frac{1}{g} \right) + O(1/(-g)^2).
\ee
This is consistent with the Gaussian approximation \eqref{GF0b}.

\section{Double scaling limits to multi-critical point}

We would like to compare the double scaling limit of the spectral curve of the perturbed model with
the SW curve of the $(A_1, A_{4k-1})$ AD theory.
At the $k$-th critical point where $t_{\pm n}=t^{(k)}_n$ and $g=1$, the spectral curve of the multi-critical matrix model degenerates to
\be
y^2 = \left(t^{(k)}_k \right)^2 \frac{(z+1)^{4k}}{z^{2k+2}}.
\ee
It has an $A_{4k-1}$ singularity at $z=-1$. We may interpret 
the perturbation as deformations of this singularity, and the double scaling limit
as a blowup at the multi-critical point.

To make the double scaling limit of the spectral curve isomorphic to the SW curve,
we add the logarithmic term to the potential as perturbation. The potential \eqref{Wz} is generalized to
\be
W(z) = \sum_{n \geq 1} \frac{1}{n} ( t_n \, z^n + t_{-n}\, z^{-n}) + t_0 \, \ln z.
\ee
For the partition function \eqref{ZG1} being well-defined, $N \, t_0$ must be an integer. We write
$t_0 = n_0/N$ with $n_0 \in \mathbb{Z}$. 

\subsection{Double scaling limit in one-cut regime}

In this subsection, we consider the case of the perturbed model remaining in the one-cut regime. 
Inclusion of the logarithmic term breaks the reality condition of the potential and causes many subtleties in the 
saddle point approach. Here we assume that $t_0$ can be analytically continued to a pure imaginary number
and the one-cut ansatze \eqref{Mz}, \eqref{EQab}, \eqref{PR2} still work after the continuation.
At the last stage of calculation, $t_0$ is continued  back to the rational number.

Then, the conditions \eqref{EQab} are modified to
\bel{pcd1a}
1 = t_0 + \sum_{n \geq 1} t_n\, v^n\, P_{n-1}(x) 
- \sum_{n \geq 1} t_{-n}\, v^{-n}\,  P_n(x),
\ee
\bel{pcd2a}
1 = \sum_{n \geq 1} t_{-n}\, v^{-n}\,  P_{n-1}(x)
 -t_0 -  \sum_{n \geq 1} t_n\, v^n\,  P_n(x).
\ee
The Laurent polynomial $M(z)$ is still given by \eqref{Mz2}, but now 
$x$ and $v$ in \eqref{Mz2} are solutions to
these modified equations \eqref{pcd1a} and \eqref{pcd2a}.

These two conditions are rewritten as
\bel{pcd1b}
2 = \sum_{n \geq 1} (t_n\, v^n + t_{-n}\, v^{-n}) \bigl( P_{n-1}(x) - P_n(x) \bigr),
\ee
\bel{pcd2b}
2\, t_0 =  -\sum_{n \geq 1} (t_n\, v^n - t_{-n}\, v^{-n}) \bigl( P_{n-1}(x) + P_n(x) \bigr).
\ee

The spectral curve for the unitary matrix model in the one-cut regime takes the form:
\bel{LPEQ}
y^2 = \widetilde{W}'(z)^2 + 4\, f(z) = \frac{1}{z^2} M(z)^2\, \sigma(z).
\ee

We are interested in the double-scaling limit of this spectral curve to the multi-critical point. Let us set\footnote{The standard ``lattice spacing'' is $a^{1/k}$ and we usually set $1/N = \kappa\, a^{2+(1/k)}.$
Here we have replaced $a^{1/k}$ by $a$ for notational simplicity.}
\bel{Ntoa}
\frac{1}{N} = \kappa\, a^{2k+1},
\ee
where $\kappa$ is a parameter with the scaling dimension $1$ and $a$ is an infinitesimal
parameter of dimension $-1/(2k+1)$.
The double-scaling limit is  taken by sending $a \rightarrow 0$ with $\kappa$ kept finite.
Note that $t_0 = n_0\, \kappa\, a^{2k+1}$.

\subsubsection{$k=1$ case}

First, we  study the $k=1$ case. Here, for simplicity, we do not turn on 
higher order terms for the potential. We consider 
\be
W(z) = t_1\, z + \frac{t_{-1}}{z} + t_0 \, \ln z,
\ee
with $t_0 = (n_0/N)$ for $n_0 \in \mathbb{Z}$.  
The Laurent polynomial \eqref{Mz2} for this potential
is given by
\be
M(z) = t_1 + \frac{t_{-1}}{v\, z}.
\ee
The conditions \eqref{pcd1b} and \eqref{pcd2b} become
\bel{k1cd1}
2 = (t_1\, v+ t_{-1}\, v^{-1}) (1-x),
\ee
\bel{k1cd2}
2\, t_0 = - ( t_1\, v - t_{-1}\, v^{-1}) (1+x).
\ee

Let us denote the perturbation of the parameters $t_{\pm 1}$ around the critical value $t^{(1)}_1 = 1/2$ by
\be
t_{\pm 1} = \frac{1}{2} + a\, t_{\pm 1, 1} + a^2\, t_{\pm 1, 2 } + \dotsm.
\ee
Also, we set $t_0 = n_0\, \kappa\, a^3$, 
\be
x = -1 + a\, x_1 + a^2\, x_2 + \dotsm, \qq
v=1+a\, v_1 + a^2\, v_2 + \dotsm.
\ee
Furthermore, we change the coordinate from $z$ to $Z$ by $z=v(-1+a\, Z)$.
Here the overall factor $v$ is introduced to simplify calculation.
At the last stage of calculation, 
it just shifts $Z$ by a constant.

Then we find
\be
\sigma(z) 
= 2\, x_1\, a + (Z^2 - 2\, x_1\, Z + 2\, x_2 + 4\, v_1\, x_1 ) a^2 + O(a^3),
\ee
\be
M(z) = \left( - \frac{1}{2} Z + t_{1,1} - t_{-1,1} + v_1 \right) a + O(a^2).
\ee
From the $O(a)$ condition of \eqref{k1cd1}, we find
$x_1 = 2 (t_{1,1} + t_{-1,1})$.
We fine-tune the perturbation by requiring $x_1=0$, i.e. $t_{1,1} = - t_{-1,1}$.
For notational simplicity, we set $t_{1,1} = (1/2) \lambda_1$.
Then, the $O(a^2)$ condition of \eqref{k1cd1} and the $O(a^3)$ condition of \eqref{k1cd2}
are respectively given by
\be
v_1^2 + 2\, v_1\, \lambda_1 - x_2 + 2 ( t_{1,2} + t_{-1,2}) = 0, \qq
( v_1 + \lambda_1) x_2 + 2\, n_0 \, \kappa= 0.
\ee
From these equations, $v_1$ is determined as
\be
v_1 = - \lambda_1 - \frac{2\, n_0\, \kappa}{x_2},
\ee
where $x_2$ is a solution to the cubic equation:
\bel{k1cubic}
x_2^3 + \bigl( \lambda_1^2 - 2 (t_{1,2} + t_{-1,2}) \bigr) x_2^2 - 4\, n_0^2\, \kappa^2 = 0.
\ee
The spectral curve $y^2 = (1/z^2) M(z)^2 \sigma(z)$ takes the form
\be
y^2 = \frac{1}{4} ( Z - 2\, \lambda_1  - 2\, v_1 )^2 (Z^2 + 2\, x_2) a^4 + O(a^5).
\ee
Note that the dependence on $(t_{1,2} + t_{-1,2})$ enters in the leading form of the spectral curve through $x_2$.
To the leading order in $a$, the combination $(t_{1,2} - t_{-1,2})$ does not appear.  
Hence for the minimal choice of perturbation, we set $t_{1,2} = t_{-1,2} = - (1/2) \mu_0$.
The minimally perturbed potential is determined as
\be
W(z) = \frac{1}{2} (1 - a^2\, \mu_0) \left( z + \frac{1}{z} \right)
+ \frac{1}{2} a\, \lambda_1\left( z - \frac{1}{z} \right) + a^3\, n_0\, \kappa \ln z.
\ee
The cubic equation \eqref{k1cubic} becomes
\bel{k1cubic2}
x_2^3 + \bigl( \lambda_1^2 + 2\, \mu_0 \bigr) x_2^2 - 4\, n_0^2\, \kappa^2 = 0.
\ee

By changing the coordinates from $(Z, y)$ to $(\tilde{z}, \tilde{y})$,
\be
Z = \tilde{z} + v_1 + \lambda_1, \qq y = \frac{1}{2}  a^2\, \tilde{y},
\ee
we obtain the double scaling limit of the spectral curve:
\bel{DSLLP}
\tilde{y}^2 = \Bigl(Z- 2(v_1+\lambda_1) \Bigr)^2(Z^2+ 2\, x_2 )
= \tilde{z}^4 + c_2\, \tilde{z}^2 + c_3 \, \tilde{z} + c_4,
\ee
where
\be
\begin{split}
c_2 &= 2\, x_2 - 2( v_1 + \lambda_1)^2=-2(\lambda_1^2+2\, \mu_0), \cr
c_3 &= - 4\, x_2 ( v_1 + \lambda_1)=8\, n_0\, \kappa, \cr
c_4 &= ( v_1 + \lambda_1)^2 \Bigl( 2\, x_2 + ( v_1+ \lambda_1)^2 \Bigr)
=( x_2 + \lambda_1^2 +2\, \mu_0)(3\, x_2 + \lambda_1^2 + 2\, \mu_0).
\end{split}
\ee
This is consistent with the result of \cite{IOY1812}: $c_2$ comes from perturbation of the couplings $t_{\pm 1}$,
$c_3$ from the coefficient of the logarithmic term, and $c_4$ from  a moduli.

The spectral curve \eqref{DSLLP} is isomorphic to the SW curve of the $(A_1, A_{3})$ AD theory. It
is degenerated. It has taken over the one-gap structure.
The discriminant $\Delta$ of the quartic polynomial $\tilde{z}^4+ c_2\, \tilde{z}^2 + c_3 \, \tilde{z} + c_4$
vanishes due to the cubic equation \eqref{k1cubic2}:
\be
\begin{split}
\Delta &=\frac{4}{27} \bigl(c_2^2+12\, c_4 \bigr)^3 
- \frac{1}{27} \bigl(2\, c_2^3 - 72\, c_2\, c_4 + 27\, c_3^2 \bigr)^2 \cr
&=256 \Bigl( x_2^3 + 2\, \tilde{\mu}_0\, x_2^2 - 4\, n_0^2\, \kappa^2 \Bigr)
\Bigl( (3\, x_2 + 2\, \tilde{\mu}_0)(3\, x_2 + 8\, \tilde{\mu}_0)^2 + 108\, n_0^2\, \kappa^2 \Bigr) = 0,
\end{split},
\ee
where $\tilde{\mu}_0:=\mu_0 +(1/2) \lambda_1^2$.

\vspace{5mm}

\noindent
\textit{Remark}. For given $t_{\pm 1}$ and $t_0$, the conditions \eqref{k1cd1} and \eqref{k1cd2}
can be rewritten as
\be
t_{1}\, v = \frac{1}{1-x} - \frac{t_0}{1+x}, \qq
t_{-1} \, v^{-1} = \frac{1}{1-x} + \frac{t_0}{1+x}.
\ee
Then $x$ is a solution to a quartic equation:
\be
t_1\,  t_{-1} (1-x)^2(1+x)^2 = (1+x)^2 - t_0^2 (1-x)^2.
\ee
By using one of the solutions for $x$,  the constant $v$ is determined as follows:
\be
v =\sqrt{ \frac{t_{-1}}{t_1}}
\left[ \frac{\dis 1 - t_0 \left( \frac{1-x}{1+x} \right)}
{\dis 1 +t_0 \left( \frac{1-x}{1+x} \right)} \right]^{1/2}.
\ee

\subsubsection{General $k$ case}

Next, we study the case of general $k$.
Let us consider the perturbation of the couplings around the $k$-th multi-critical fixed point
\be
t_n = t^{(k)}_n + \delta t_n, \qq
t_{-n} = t^{(k)}_{n} + \delta t_{-n}, \qq
(n \geq 1),
\ee
and $t_0 = \delta t_0$. We write the potential as
\be
W(z) = W^{(k)}_0(z) + \delta W(z),
\ee
where
\be
W_0^{(k)}(z) = \sum_{n=1}^k \frac{t^{(k)}_n}{n}( z^n + z^{-n}), \qq
\delta W(z) = \sum_{n \geq 1} \frac{1}{n} ( \delta t_n \, z^n + \delta t_{-n}\, z^{-n})
+ \delta t_0 \ln z.
\ee
We also introduce the ``dressed perturbation''  $\delta \tilde{t}_{\pm n}$ by
\be
v^n\, t_n = t^{(k)}_n + \delta \tilde{t}_n, \qq
v^{-n}\, t_{-n} = t^{(k)}_n + \delta \tilde{t}_{-n}, \qq
(n \geq 1).
\ee
Let us denote the perturbative expansion of various objects around the multi-critical point by
\be
x = -1 + \delta x, \qq \delta x = a\, x_1+ a^2\, x_2 + a^3\, x_3 + a^4\, x_4+\dotsm,
\ee
\be
z =v( -1  +a Z),
\qq
v=1+ a\, v_1 + a^2\, v_2 + \dotsm,
\ee
\be
\delta t_{n} = a \, t_{n, 1} + a^2\, t_{n, 2} + \dotsm, \qq (n \neq 0),
\qq
\delta t_0 = n_0\, \kappa\, a^{2k+1}.
\ee
Note that 
the quadratic differential $y^2\, \de z^2$ has two poles of degree $2k+2$ at $z=0$ and at $z=\infty$.
In the $Z$ coordinate, these poles locate at $Z=1/a$ and $Z=\infty$, respectively.
In the $a \rightarrow 0$ limit, these two poles merge into a pole at $Z=\infty$.

The dimensions of these parameters are assigned as follows
\be
[ x_i ] = [v_i]=[ t_{n,i}] =\frac{i}{2k+1}, \qq
[ Z] = \frac{1}{2k+1}.
\ee
The conditions \eqref{pcd1b} and \eqref{pcd2b} can be written as
\bel{pcd1r}
\left( \frac{\delta x}{2} \right)^k 
+ \sum_{m \geq 0} 
\left( \frac{\delta x}{2} \right)^m \delta \mathcal{A}^{(m)}=0,
\ee
\bel{pcd2r}
\delta t_0 + \sum_{m \geq 1} m \left( \frac{\delta x}{2} \right)^m
\delta \mathcal{B}^{(m)}=0,
\ee
where
\bel{dAm}
\delta \mathcal{A}^{(m)}:=\sum_{n \geq \mathrm{max}(1,m)} (-1)^{n-m} 
\frac{n\, \Gamma(n+m)}{(m!)^2\, \Gamma(n-m+1)} ( \delta \tilde{t}_n + \delta \tilde{t}_{-n}),
\ee
\bel{dBm}
\delta \mathcal{B}^{(m)}
:=\sum_{n \geq m} (-1)^{n-m}
\frac{\Gamma(n+m)}{(m!)^2\, \Gamma(n-m+1)} 
 ( \delta \tilde{t}_n - \delta \tilde{t}_{-n}).
\ee
Let us denote the expansion of the dressed perturbation parameters by
\be
\delta \tilde{t}_{\pm n} = \sum_{r \geq 1} a^r\, \tilde{t}_{\pm n, r}.
\ee
We also denote the $a$-expansion coefficients of $\delta \mathcal{A}^{(m)}$ and
$\delta \mathcal{B}^{(m)}$ by
\be
\delta \mathcal{A}^{(m)}= \sum_{r \geq 1} a^r\, \mathcal{A}^{(m)}_r,
\qq
\delta \mathcal{B}^{(m)}
=\sum_{r \geq 1} a^r\, \mathcal{B}^{(m)}_r,
\ee
where
\be
\mathcal{A}^{(m)}_r:=\sum_{n \geq \mathrm{max}(1,m)} (-1)^{n-m} 
\frac{n\, \Gamma(n+m)}{(m!)^2\, \Gamma(n-m+1)} ( \tilde{t}_{n,r} +\tilde{t}_{-n,r}),
\ee
\be
\mathcal{B}^{(m)}_r
:=\sum_{n \geq m} (-1)^{n-m}
\frac{\Gamma(n+m)}{(m!)^2\, \Gamma(n-m+1)} 
 ( \tilde{t}_{n,r} - \tilde{t}_{-n,r}).
\ee

In the double scaling limit, the coupling constants should be fine-tuned such that
zero points of the spectral curve \eqref{LPEQ} approach to the critical point.
The double scaling limit of the spectral curve 
is isomorphic to the SW curve of the $(A_1, A_{4k-1})$ AD theory.

The coupling constants should be fine-tuned by imposing the conditions:
\bel{condcc}
M(z) = O(a^{2k-1}), \qq \sigma(z) =O(a^2).
\ee
Let $M(z) = a \, M_1 + a^2\, M_2 + \dotsm$, and
$\sigma(z) = a\, \sigma_1 + a^2 \, \sigma_2+O(a^3)$. 
Here $\sigma_1 = 2\, x_1$.
Then \eqref{condcc} are equivalent to
$M_1 = M_2 = \dotsm = M_{2k-2} = 0$ and  $x_1=0$.
Also taking into account of \eqref{pcd1r} up to $O(a^{2k-1})$ and
\eqref{pcd2r} up to $O(a^{2k})$, 
we obtain the constraints:
\bel{cAr}
\mathcal{A}^{(m)}_r=0, \qq
(0 \leq m \leq k-1, \ 
1 \leq r \leq 2\, k-1-2\, m ),
\ee
\bel{cBr}
\mathcal{B}^{(m)}_r =0, \qq
(1 \leq m \leq k-1, \  1 \leq r \leq 2\, k - 2\, m).
\ee
Equivalently, these constraints are stated as follows:
\begin{align} \label{leadingAB}
\delta \mathcal{A}^{(m)} &= a^{2k-2m}\, \mathcal{A}^{(m)}_{2k-2m} + O(a^{2k-2m+1}),&
(0 \leq & m \leq k-1),
\cr
\delta \mathcal{B}^{(m)} &= a^{2k-2m+1} \, \mathcal{B}^{(m)}_{2k-2m+1} + O(a^{2k-2m+2}),&
(1 \leq & m \leq k).
\end{align}

For simplicity, we further assume that $\delta t_{\pm n}=0$ for $n>k$.
We turn on $\delta t_{\pm n}$ only for $1 \leq n \leq k$.
Then, the constraints for $\mathcal{A}^{(m)}_r$ and $\mathcal{B}^{(m)}_r$
determine the leading terms of $t_{n,r} \pm t_{-n, r}$ in the $a$-expansion. 
We find that the minimally perturbed potential takes the form
\be
\begin{split}
& \sum_{n=1}^k \frac{1}{n} ( \delta t_n \, z^n +
\delta t_{-n}\, z^{-n}) \cr
&= \sum_{m=0}^{k-1} a^{2k-2m} \mu_m\,
\Bigl( W^{(m)}_0(z) - W^{(k)}_0(z) \Bigr)
+ \sum_{m=1}^k a^{2k-2m+1} \lambda_{m}
\left( z \frac{\de}{\de z} W^{(m)}_0(z) \right).
\end{split}
\ee
Here $\mu_m$ and $\lambda_m$ are free parameters with dimensions
\be
[ \mu_m] = \frac{2k-2m}{2k+1}, \qq
[ \lambda_m] = \frac{2k-2m+1}{2k+1}.
\ee
In terms of $\mu_n$ and $\lambda_n$,
the lowest non-trivial coefficients \eqref{leadingAB} of $\delta \mathcal{A}^{(m)}$ and $\delta \mathcal{B}^{(m)}$ are
given by
\be
\mathcal{A}^{(m)}_{2k-2m}
=\sum_{n=m}^k (-1)^{n-m}
\frac{(n!)^2}{(m!)^2 \, (2(n-m))!} v_1^{2(n-m)-1}
\Bigl(v_1\, \mu_n + 2(n-m) \lambda_n \Bigr),
\ee
\be
\mathcal{B}^{(m)}_{2k-2m+1}
= \sum_{n=m}^k (-1)^{n-m}
\frac{(n!)^2}{(m!)^2\, (2(n-m)+1)!} v_1^{2(n-m)} \Bigl(v_1\, \mu_n + \bigl(2(n-m)+1 \bigr) \lambda_n \Bigr),
\ee
with $\mu_k=1$. 
Note that only $v_1$ enters in these coefficients. 
The $O(a^{2k})$ condition of \eqref{pcd1r} and the $O(a^{2k+1})$ condition of \eqref{pcd2r}
are given respectively by
\be
\left( \frac{x_2}{2} \right)^k
+ \sum_{m=0}^{k-1} \left( \frac{x_2}{2} \right)^m \mathcal{A}^{(m)}_{2k-2m} = 0,
\qq
n_0\, \kappa + \sum_{m=1}^k m \left( \frac{x_2}{2} \right)^m
\mathcal{B}^{(m)}_{2k-2m+1} = 0.
\ee
This system of two algebraic equations determines $x_2$ and $v_1$ in terms of $\mu_n$ and $\lambda_n$.

The spectral curve \eqref{LPEQ} takes the form
\be
y^2 = a^{4k} \bigl( M_{2k-1} \bigr)^2\, \sigma_2 + O(a^{4k+1}),
\ee
with $\sigma_2 = Z^2 +2\, x_2$, and
\be
\begin{split}
M_{2k-1} 
&=t^{(k)}_k \sum_{\ell=1}^{k} (-1)^{\ell}\,
\frac{(1/2)_{k-\ell}}{(1)_{k-\ell}} \,(2 x_2)^{k-\ell} \, Z^{2\ell-1} 
+ \sum_{\ell=1}^{k-1}
\sum_{r=\ell+1}^{k} 
(-1)^{\ell} 
\, \mathcal{C}^{(+)}_{k-r,\ell} 
\, \mathcal{A}^{(k+\ell-r)}_{2r-2\ell}
\, x_2^{k-r}
\, Z^{2\ell-1} \cr
& + \sum_{\ell=0}^{k-1} \sum_{r=\ell+1}^{k}  
(-1)^{\ell}   \mathcal{C}^{(-)}_{k-r,\ell} 
\, \mathcal{B}^{(k+\ell-r+1)}_{2r-2\ell-1}
\, x_2^{k-r} 
\, Z^{2\ell},
\end{split}
\ee
where
\be
\mathcal{C}_{r, \ell}^{(+)}:= \frac{(r+\ell)!}{2^{r+2\ell}\, r!}
\frac{\Gamma(r+1/2)}{\Gamma(r+\ell+1/2)},
\ee
\be
\mathcal{C}_{r, \ell}^{(-)}:= (r+\ell+1)^2 \frac{(r+\ell)!}{2^{r+2\ell+1}\, r!}
\frac{\Gamma(r+1/2)}{\Gamma(r+\ell+3/2)}.
\ee

If we set
\be
M_{2k-1} = (-1)^k\, t^{(k)}_k \mathscr{M}_{2k-1}(Z),
\ee
then $\mathscr{M}_{2k-1}$ is a monic polynomial of $Z$ with degree $2k-1$:
\be
\mathscr{M}_{2k-1} = Z^{2k-1} - 2k\, \mathcal{B}^{(k)}_1\, Z^{2k-2} + \dotsm.
\ee
Now by setting
$y =t^{(k)}_k  \, a^{2k}\,  \tilde{y}$ with  $[ \tilde{y} ] = 2k/(2k+1)$,
we obtain the double-scaling limit of the spectral curve \eqref{LPEQ}:
\bel{DSLLPEQ}
\tilde{y}^2 = \bigl( \mathscr{M}_{2k-1}(Z) \bigr)^2 (Z^2 +2\, x_2).
\ee
This curve has taken over the one-gap structure.
Shifting the variable $Z$ by
$Z = \tilde{z} + \mathcal{B}^{(k)}_1$ with 
$[ \tilde{z}] = 1/(2k+1)$,
the spectral curve \eqref{DSLLPEQ} takes the form of the SW curve of the $(A_1, A_{4k-1})$ AD theory:
\bel{SWC}
\tilde{y}^2 = \tilde{z}^{4k} + c_2\, \tilde{z}^{4k-2} + c_3\, \tilde{z}^{4k-3}
+ \dotsm + c_{4k-1} \tilde{z} +c_{4k}.
\ee
Here $[c_i]=i/(2k+1)$.
Note that $[ \tilde{y}] + [ \tilde{z}] =1$.
The quadratic differential $\tilde{y}^2\, \de \tilde{z}^2$ has a pole of degree $4k+4$
at $\tilde{z}=\infty$.
.

\vspace{5mm}

\noindent
\textit{Remark}. By using the one-cut solution, 
we have shown that  $W_0^{(m)}(z) - W_0^{(k)}(z)$ and $z (\de W_0^{(m)}(z)/\de z)$
form a basis of perturbation. On the other hand, 
the analysis by the string equation in \cite{IY2103} implies that 
$W_0^{(m)}(z)$ and their derivatives $z (\de W_0^{(m)}(z)/\de z)$
form another basis.
Relation between these two bases may be understood as follows.
Let us rewrite the ``even part'' of the potential as
\be
\begin{split}
& W_0^{(k)}(z) + 
\sum_{m=0}^{k-1} a^{2k-2m} \mu_m\,
\Bigl( W^{(m)}_0(z) - W^{(k)}_0(z) \Bigr) \cr
&=g\, W_0^{(k)}(z)+ \sum_{m=1}^{k-1} a^{2k-2m} \mu_m\, W^{(m)}_0(z),
\end{split}
\ee
where
\be
g = 1 - \sum_{m=0}^{k-1} a^{2k-2m} \mu_m.
\ee
We interpret the appearance of $g$ as a consequence of ``renormalization'' of the unperturbed potential $W^{(k)}_0(z)$.

\subsection{Double scaling limit in multi-cut regime}

In this subsection, we consider the case of the perturbed model transferred to the multi-cut regime
with maximal number of cuts.
Let us consider a degree $k$ potential with a logarithmic term:
\be
W(z) = \sum_{n=1}^k \frac{1}{n} ( t_n\, z^n + t_{-n}\, z^{-n}) + t_0\, \ln z.
\ee
The spectral curve for this potential in the $2k$-cut regime takes the form
\bel{SC2k}
y^2 = \widetilde{W}'(z)^2 + 4\, f(z)
= \frac{t_k{}^2}{z^{2k+2}} \prod_{i=1}^{2k} ( z - \alpha_i )(z- \beta_i).
\ee
The zero points of the spectral curve, $\alpha_i$ and $\beta_i$, are determined from the
coupling constants $t_{\pm n}$, $t_0$ and filling fractions $\nu_i=N_i/N$ with
$\sum_i \nu_i = 1$.
Here $N_i$ denotes the number of eigenvalues localized near the $i$-th extremum of the potential.
We assume that all the zero points $\alpha_i$, $\beta_j$ are different from each other.

We set $(1/N) = \kappa\, a^{2k+1}$, 
\be
t_{\pm n} = t^{(k)}_n + O(a), \qq
t_0 = n_0 \, \kappa\, a^{2k+1}, 
\qq
z = -1 + a\, Z, \qq
y =t^{(k)}_k \, a^{2k}\, \tilde{y}.
\ee
In the double scaling limit, 
the couplings $t_{\pm n}$ should be fine-tuned by requiring
\be
\alpha_i = -1 + a\, \widetilde{\alpha}_i + O(a^2), \qq
\beta_i = -1 + a\, \widetilde{\beta}_i + O(a^2).
\ee
Then, the double scaling limit of the spectral curve \eqref{SC2k} is given by
\be
\tilde{y}^2 = \prod_{i=1}^{2k} (Z- \widetilde{\alpha}_i)(Z - \widetilde{\beta}_i).
\ee
By a constant shift of $Z$, this takes the form of the SW curve
of the $(A_1, A_{4k-1})$ AD theory \eqref{SWC}.
We expect that the scaling behavior of the fine-tuned couplings is essentially the same as that
in the one-cut regime. A detailed study will be given elsewhere.

\section{Discussion}

We have studied the unitary matrix model with the multi-critical coupling constants, and
determined the eigenvalue density, the planar Wilson loops, and the planar free energy in the one-gap phase.
Furthermore, the double scaling limit of the matrix model spectral curve has been examined.
It is isomorphic to the SW curve of the $(A_1, A_{4k-1})$ AD theory.

The planar resolvent in the ungapped phase is given by
\[
\omega^{(\mathrm{ungapped})}(z) = \frac{1}{z} + \sum_{n=1}^k \frac{t^{(k)}_n g}{z^{n+1}},
\]
while that in the gapped phase Ia is determined as \eqref{PRIa}.
The third term in the right-hand side of \eqref{PRIa} expresses
deviation from the form in the ungapped phase. We don't know why it can be written in 
terms of the hypergeometric function of
\[
\left( 1 - \frac{1}{g} \right)^{1/k} \frac{4\, z}{(z+1)^2}.
\]
 
The unperturbed $k$-th multi-critical unitary matrix models contain only one free parameter $g$.
They have various common features with the GWW model ($k=1$ case).
For the GWW model, finite $N$ corrections and instanton corrections are studied in \cite{gol80,mar0805,oku1705,AB1707,jha2003}. 
Also, the contribution from the complex saddles are investigated in \cite{BDV1512}.
In this paper, we have considered the planar contributions only.
It may be interesting to investigate the finite $N$ and instanton corrections for these unperturbed models.

It is known that the unitary matrix models with general coupling have many phases 
\cite{AMM1610,CDD1708,RT2007,rus2010,ST2102}.
We have considered the multi-critical models with real coupling $g$, and showed that they have three phases.
If we extend the coupling $g$ to the complex region, the models are no longer unitary, and become
holomorphic matrix models \cite{laz0303}. The holomorphic models would have more rich phase structures.
It will be challenging to study them.

\section*{Acknowledgments}

The author would like to thank Hiroshi Itoyama and Katsuya Yano for useful discussions.

\appendix

\section{Some integral formula}

The integral formula \eqref{IF1} and the $n=1$ case of \eqref{IF2} 
are used to evaluate the planar free energy $F_0$ \eqref{F0rho} in the one-gap phase.
\bel{IF1}
\int_0^1 \de x\, x^{2r} \sqrt{1-x^2} = \frac{\sqrt{\pi}\, \Gamma(r+1/2)}{4\, \Gamma(r+2)},
\ee
\bel{IF2}
\int_0^1 \de x\, x^{2m} \, (1-x^2)^{n-(1/2)} \ln(2x)
= \frac{\Gamma(m+1/2) \Gamma(n+1/2)}{4\, \Gamma(m+n+1)}
\Bigl[ \psi(m+1/2) - \psi(n+m+1) + 2\, \ln 2 \Bigr].
\ee


\end{document}